\documentclass[namedreferences]{kluwer}
\usepackage{epsfig}
\def\la{\;\raise0.6mm\hbox{$<$}\hspace{-2.8mm}
        \raise-0.8mm\hbox{\small $\sim$}\;}

\begin{document}

\begin{article}

\begin{opening}

\title{GAS DYNAMICS IN THE GALACTIC BAR REGION
       FROM N-BODY AND SPH SIMULATIONS}
\author{R. \surname{FUX}}
\institute{Geneva Observatory,\\
           Ch. des Maillettes 51, CH-1290 Sauverny, Switzerland}
\runningtitle{GAS DYNAMICS IN THE GALACTIC BAR REGION}
\runningauthor{R. FUX}
\begin{abstract}
Self-consistent hybrid $N$-body and SPH simulations are used to give a
new and coherent interpretation of the main features standing out from
the HI and CO longitude-velocity observations within the Galactic bar.
In particular, the traces of the gas associated to the Milky Way's
dustlanes can be reliably identified and the 3-kpc arm appears as a
gaseous stream rather than a density wave. The bar and the gaseous
nuclear ring in the simulations undergo synchronised and long lived
oscillations around the centre of mass, and the gas flow always
remains non-stationary, suggesting a transient nature of the observed
gas kinematics.
\end{abstract}

\end{opening}

\section{Introduction}
%%%%%%%%%%%%%%%%%%%%%%
During the last decade, there has been accumulating evidence that the
Milky Way is a barred galaxy (e.g. Gerhard~1999 for a recent review),
confirming the early presumption of de Vaucouleurs~(1964) based on
the observed gas kinematics. In particular, the HI and CO
longitude-velocity ($\ell-V$) distributions betray large non-circular
motions near the Galactic centre which have been attributed to gas
moving on non self-intersecting closed $x_1$ orbits of the barred
Galactic potential. The direct detection of the bar has motivated
several recent gas dynamical studies (e.g. Wada et al. 1994; Englmaier
\& Gerhard 1999; Weiner \& Sellwood~1999).
\par
\begin{figure}[t!]
\centerline{\epsfig{file=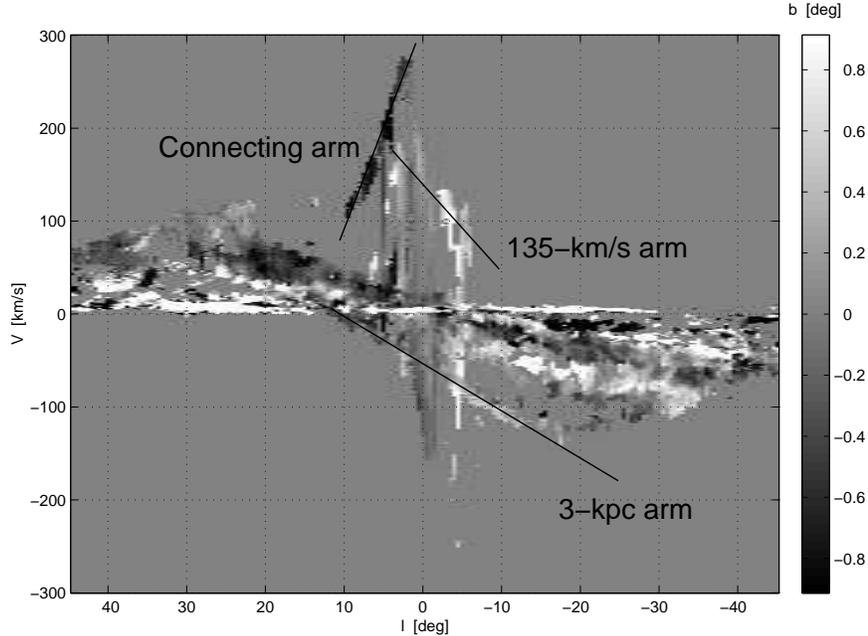,width=11.5cm}}
\caption{Emission weighted mean latitude of $^{12}$CO in the $\ell-V$
         plane. Dark and white indicate respectively gas below ($b<0$)
         and above ($b>0$) the Galactic plane. The data are from Dame
         et al. 1987.}
\label{lvb}
\end{figure}

 Fux~(1999) has tried to give a new and coherent interpretation of
the so far poorly understood dominant features seen in the gaseous
$\ell-V$ diagrams within the bar region (see Figure~\ref{lvb}), based
on high resolution $N$-body and SPH simulations and on observed
properties in external barred galaxies. The present contribution gives
a brief summary of the main results and some complements.

\section{Simulations}
%%%%%%%%%%%%%%%%%%%%%
The stellar and gas dynamics is modelled in a self-consistent way by
the $N$-body and SPH techniques, starting from plausible bar unstable
axisymmetric initial conditions. The simulations are divided into two
parts. First we run only the collisionless stellar and dark components
for 5 Gyr, keeping the gas fixed (simulation lxx), and then the gas is
softly released at two different times, $t=1800$ and $2400$~Myr, with
a sound speed $c_s=10$~km\,s$^{-1}$ (simulations l10 and l10'
respectively). The total number of particles is $\sim 4\times 10^6$,
with 150\,000 gas particles, and the simulations are evolved without
imposed symmetries. Some animations of simulation $l10$, including
live $\ell-V$ diagrams, are available at
{\tt http://obswww.unige.ch/\~{}fux}.
\par The stellar bar forms at around $t=1.2$~Gyr, well before
releasing the gas, and its pattern speed naturally adjusts to real
physical constraints. Its projected surface mass density gives a good
representation of the deredened COBE/DIRBE K-band map for realistic
viewing points relative to the bar. One of the most relevant dynamical
properties of the bar is its offcentring on Gyr timescales with
respect to the centre of mass of the global mass distribution (see
Figure~9 in Fux~1999). The displacement can reach several 100~pc and
the revolution frequency of the density centre amounts to
$20-30$~km\,s$^{-1}$\,kpc$^{-1}$.
\begin{figure}[t!]
\hspace*{.1cm}\epsfig{file=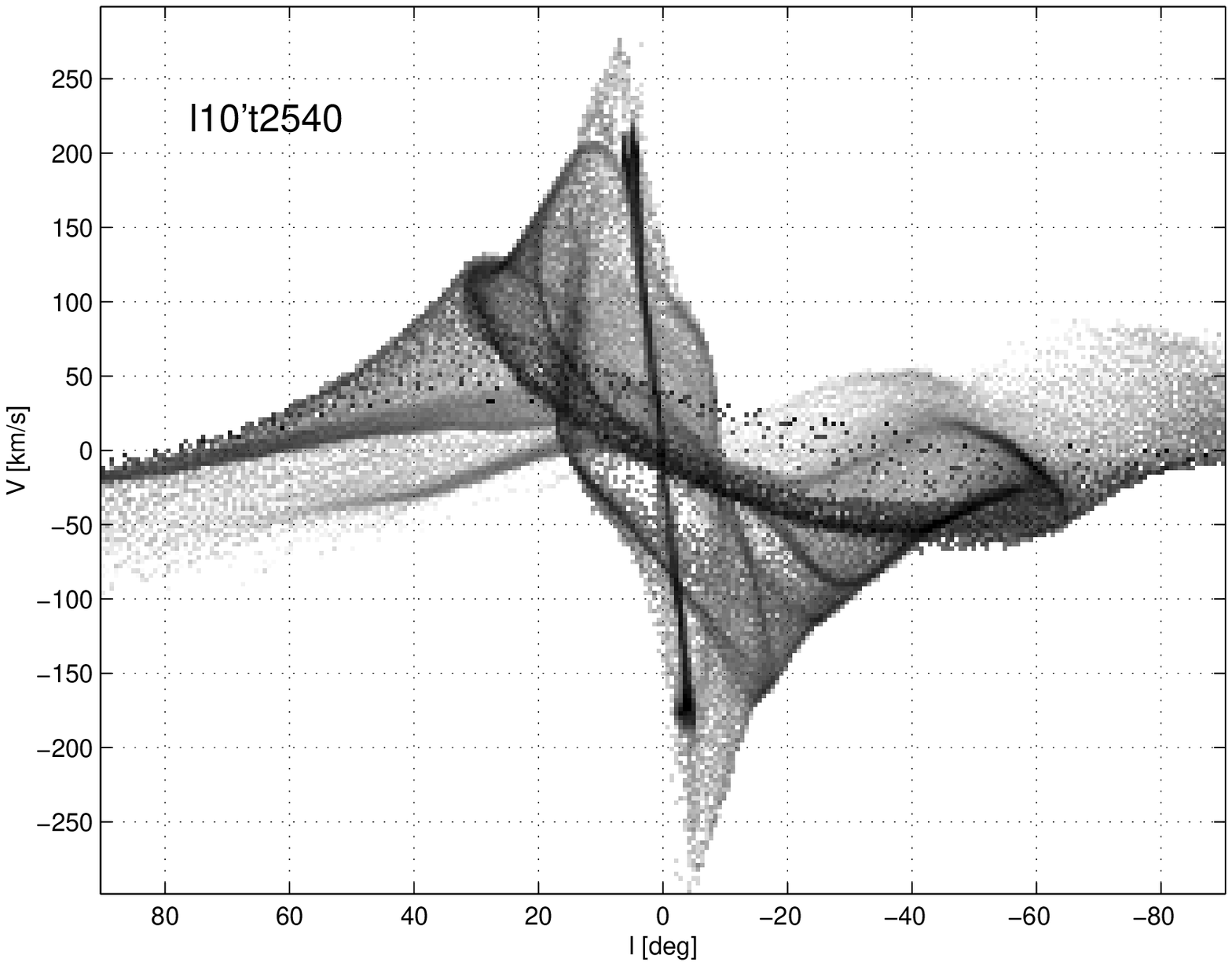,width=6.05cm}\vspace*{-4.78cm}\\
\hspace*{6.55cm}\epsfig{file=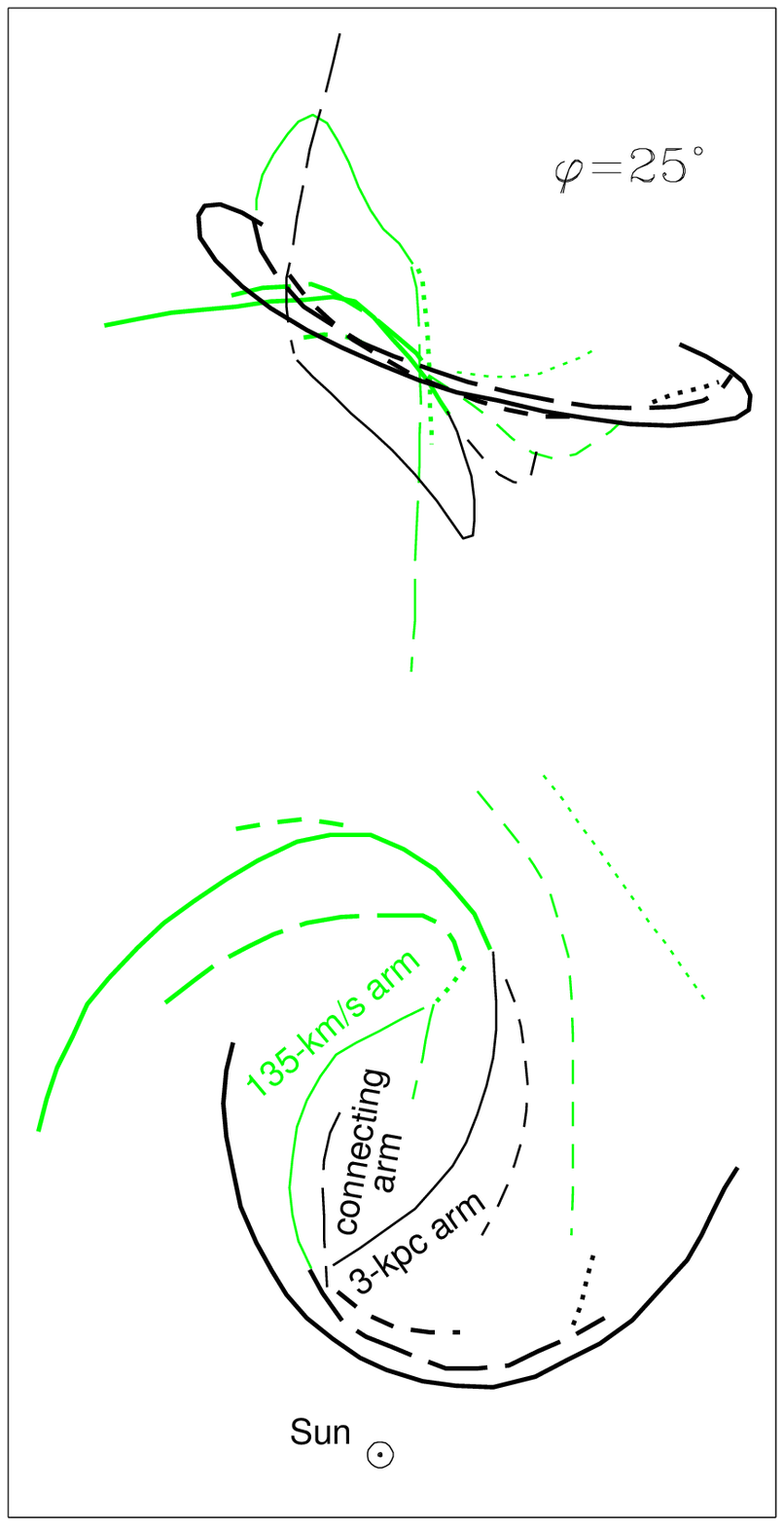,width=5.1cm}\vspace*{-4.97cm}\\
\hspace*{.335cm}\epsfig{file=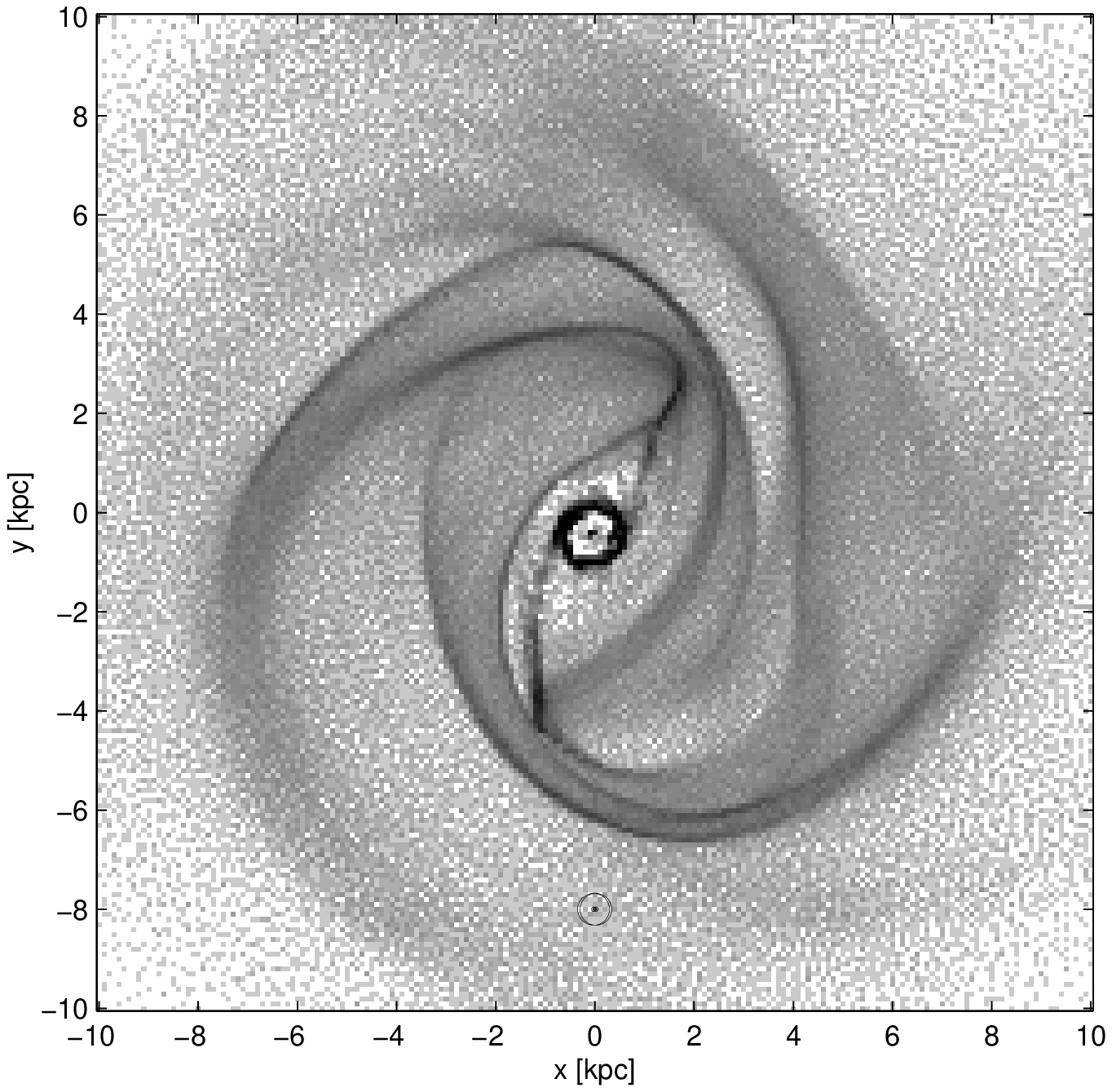,width=6.05cm}
\caption{Model l10't2540, selected as one of the models reproducing
best most of the observed features in the region
$|\ell|\la 35^{\circ}$. {\it Top left}: $\ell-V$ diagram based on the
gaseous particles within $|b|<2^{\circ}$. {\it Bottom left}: face-on
projection of the gas distribution, with the location of the observer
indicated by the $\odot$ symbol. The bar inclination angle is
$25^{\circ}$ and the corotation radius 0.55 relative to the Sun's
galactocentric distance. {\it Right}: link between the spiral arms and
their $\ell-V$ traces.}
\label{models}
\end{figure}
\par The gas flow in the simulations, which take into account the gas
self-gravity and the gravitational interaction of the gas with the
stellar arms, is non-stationary and affected by substantial
asymmetries. Beside the nuclear ring, discussed in the next section,
the gaseous structures in the bar region may be classified according
to two types (see frame $t=2066$~Myr in Figure~\ref{traj} for an ideal
case):
\begin{itemize}
\item The {\it axis shocks}, which connect the ends of the bar to the
nuclear ring and are characterised by very strong velocity gradients.
\item The {\it lateral arms}, which roughly join the bar ends avoiding
the nuclear ring by a large bow.
\end{itemize}
\par At specific times and for a location of the observer such that
the corotation radius of the bar is $4.0-4.5$~kpc and the bar
inclination angle $\varphi\sim 25^{\circ}$, the simulations reveal
models which compare fairly well with observations (see
Figure~\ref{models} for one of our favorite models).~These models
allow an original interpretation of the main HI and CO $\ell-V$
features in terms of axis shocks and lateral arms, which is further
developed in Sections~\ref{shock} and~\ref{lateral}.

\section{Nuclear ring/disc}
%%%%%%%%%%%%%%%%%%%%%%%%%%%
%
\begin{figure}[t!]
\centerline{\epsfig{file=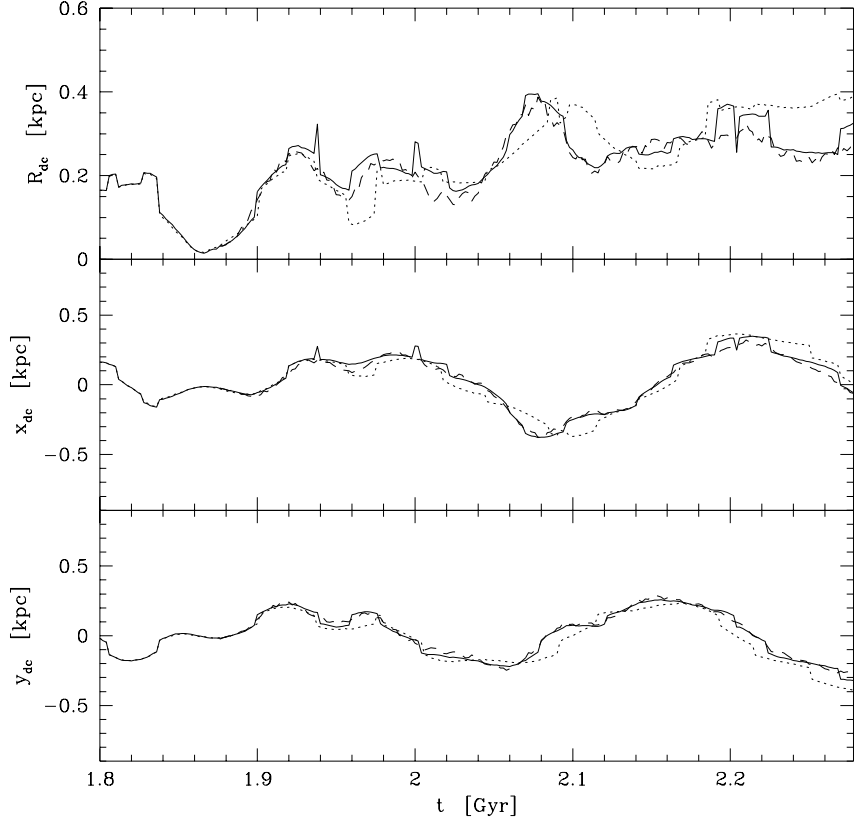,width=11.3cm}}
\caption{Radial, $x$- and $y$-displacements of the stellar density
centre (solid lines) and of the gaseous nuclear ring (dashed lines) in
the live gas simulation l10. The dotted lines indicate the
displacement of the stellar density centre in the fixed gas
simulation~lxx.}
\label{delay}
\end{figure}
The mass distribution in our simulations involves a single inner
Lindblad resonance, at $R_{\rm ILR}=1.7$~kpc and $1.6$~kpc from the
mass density centre in the scaled models l10t2066 and l10't2540
respectively. This is the radius below which the anti-bar $x_2$ orbits
exists and a nuclear ring of such orbits rapidly concentrates at
$R\la R_{\rm ILR}/2$. The nuclear ring produces a marked dip in the
$\Omega-\kappa/2$ curve, but which never reaches a value below the bar
pattern speed $\Omega_{\rm P}$, as expected for the ring to be stable.
The size of the ring is set by the geometry of the axis shocks: the
energy dissipation when the gas of these shocks hits the ring prevents
the nuclear ring to extend beyond the central offset of the shocks.
One shortcoming of our models is that the nuclear ring lies beyond the
dense nuclear molecular gas observed within $1.5^{\circ}$ from the
Galactic centre, corresponding to $R\la 200$~pc.
\par The nuclear ring closely follows the density centre displacement
of the stellar bar, apparently without a significant time delay
(Figure~\ref{delay}). The cause of the offcentring is not very clear
but it is certainly not induced by the gas component alone, because
the bar leaves the centre of mass before the gas has been released.
Moreover, a comparison between the live (l10) and fixed (lxx) gas
simulations reveals only small differences in the trajectory of the
stellar density centres, despite the fact that the bar rotates faster
in the former case.

\section{Connecting arm}
%%%%%%%%%%%%%%%%%%%%%%%%
\label{shock}
The connecting arm (Figure~\ref{lvb}) has not been given much
attention since the direct detections of the Galactic bar by Blitz \&
Spergel (1991) and others. A remarkable success of our simulations is
that some snapshots can reproduce very precisely this $\ell-V$ feature
(see Figure~\ref{models} and $t=2066$~Myr in Figure~\ref{traj}).
\par The connecting arm obviously corresponds to an axis shock. Most
high resolution hydro simulations in rotating barred potentials
display such shocks (e.g. Athanassoula 1992) and they are believed to
be related to the prominent dustlanes observed in many early-type
external barred galaxies. These dustlanes indeed seem to coincide with
gaslanes and velocity field measurements reveal velocity changes of
order 200~km\,s$^{-1}$ across them (Reynaud \& Downes 1998; Laine et
al. 1999). Thus, according to our models, the connecting arm is the
gaseous trace of the near-side branch of the Milky Way's dustlanes.
The far-side branch is seen in Figure~\ref{lvb} as a vertical feature
near $\ell=-4^{\circ}$. The same figure also reveals that the
near-side branch of these dustlanes lies below ($b<0$) the Galactic
plane and the far-side branch above it. 
\par Velocity elongated features like the one at
$\ell\approx 5.5^{\circ}$ could be gas lumps which are just about to
cross the near-side dustlane, undergoing the strong velocity gradient
of the associated shock.

\section{3-kpc arm}
%%%%%%%%%%%%%%%%%%%
\label{lateral}
\begin{figure}[p!]
\centerline{\epsfig{file=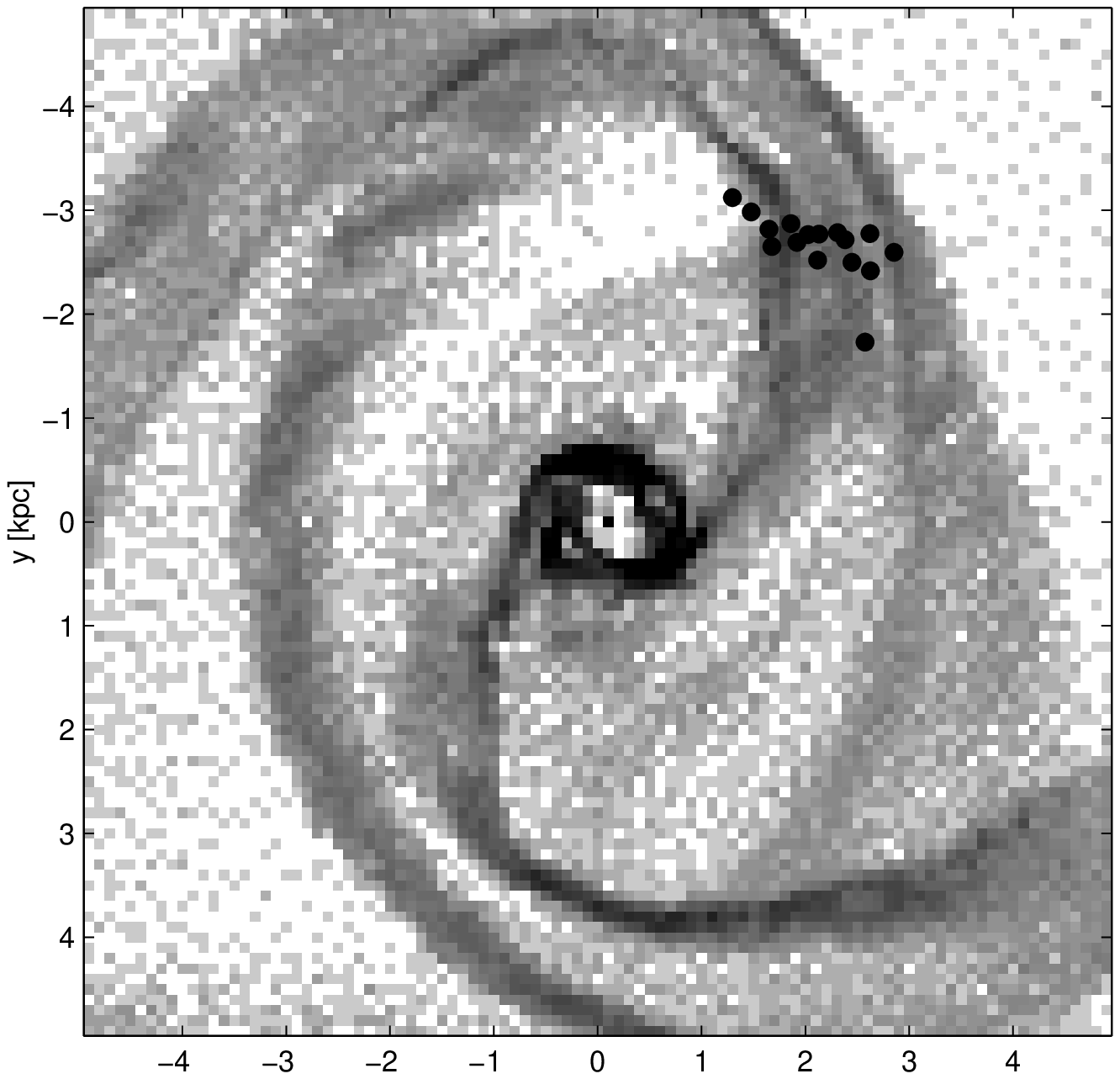,width=5.5cm}
            \epsfig{file=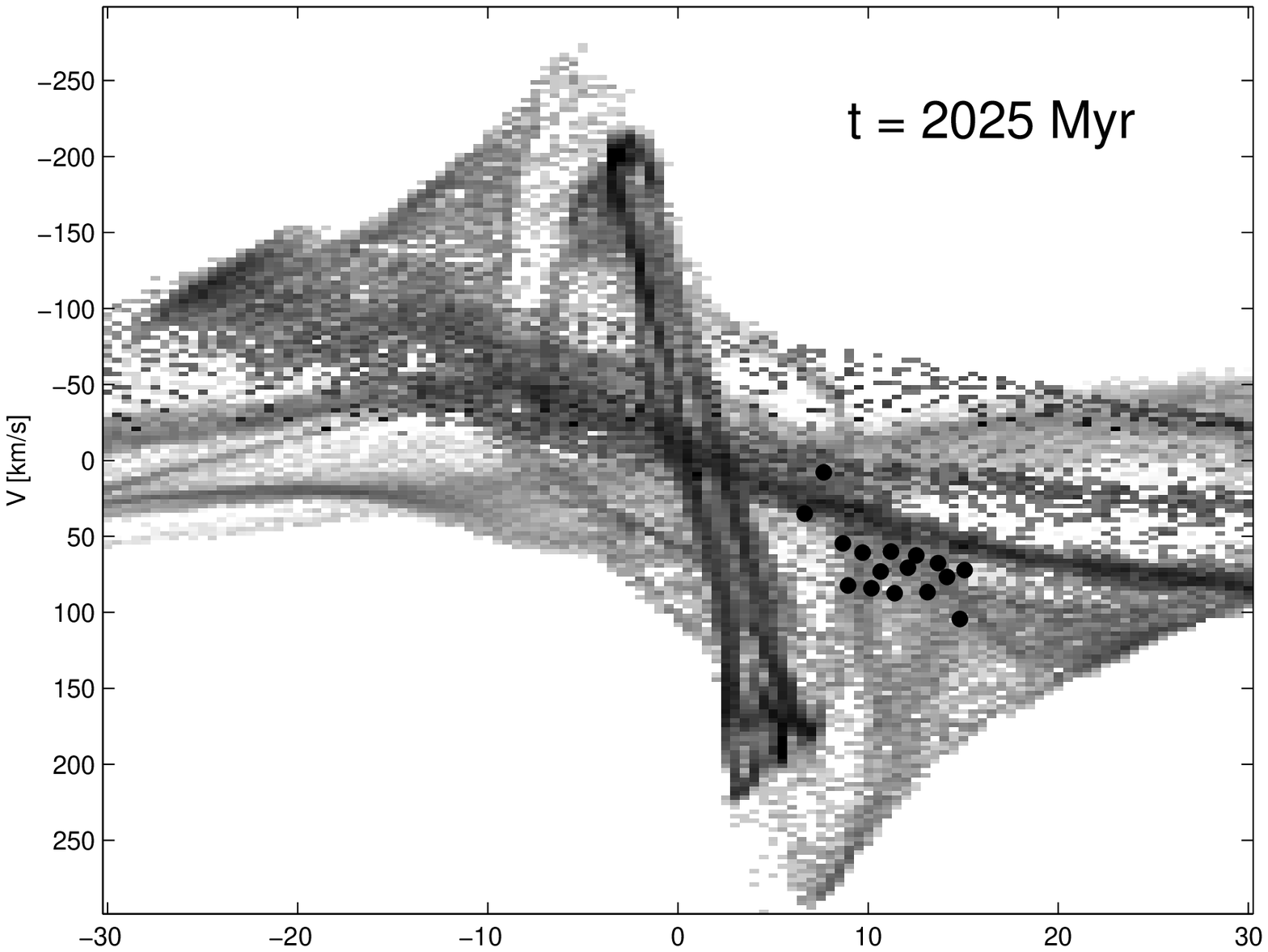,width=6.025cm}}
\centerline{\epsfig{file=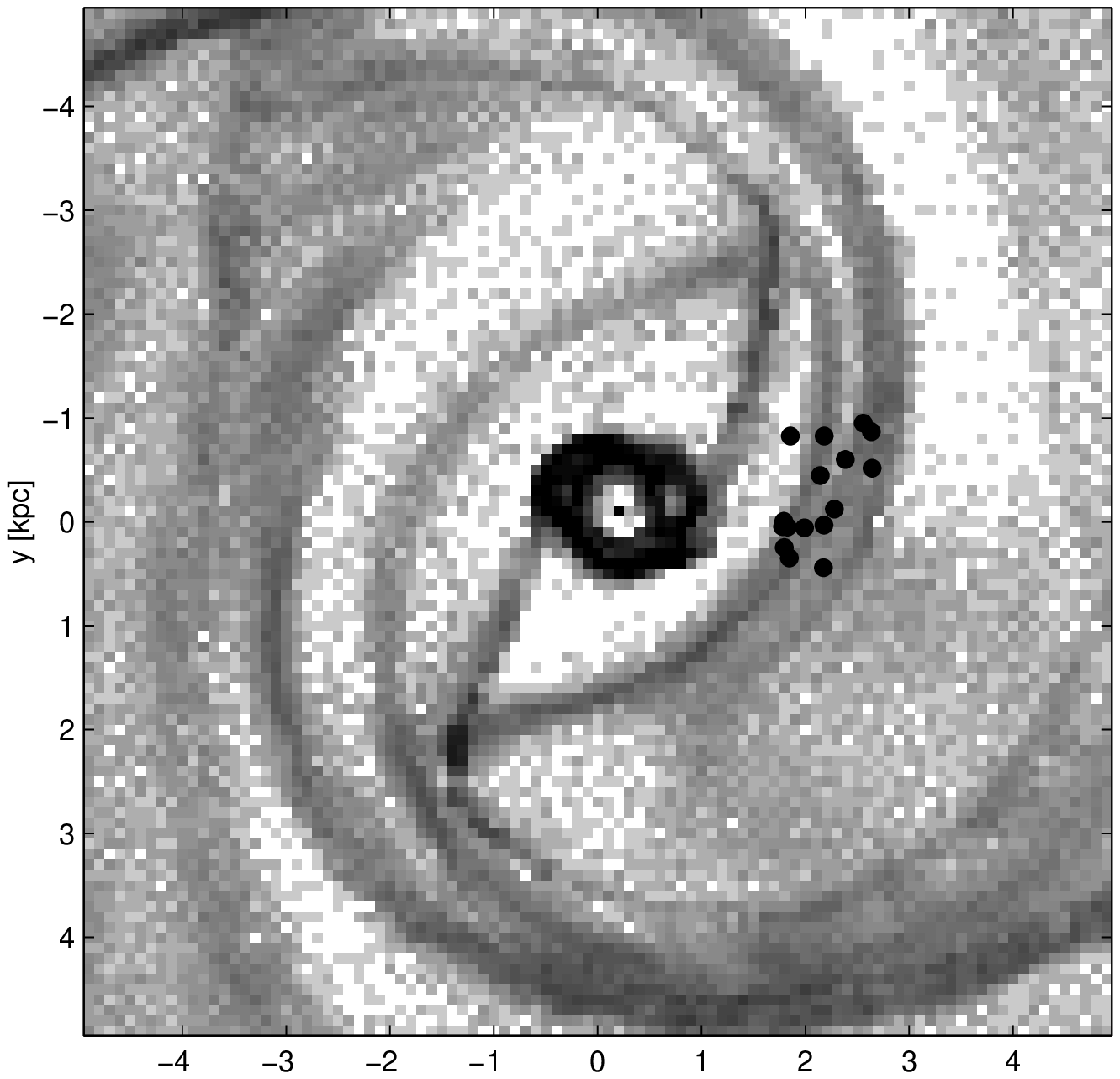,width=5.5cm}
            \epsfig{file=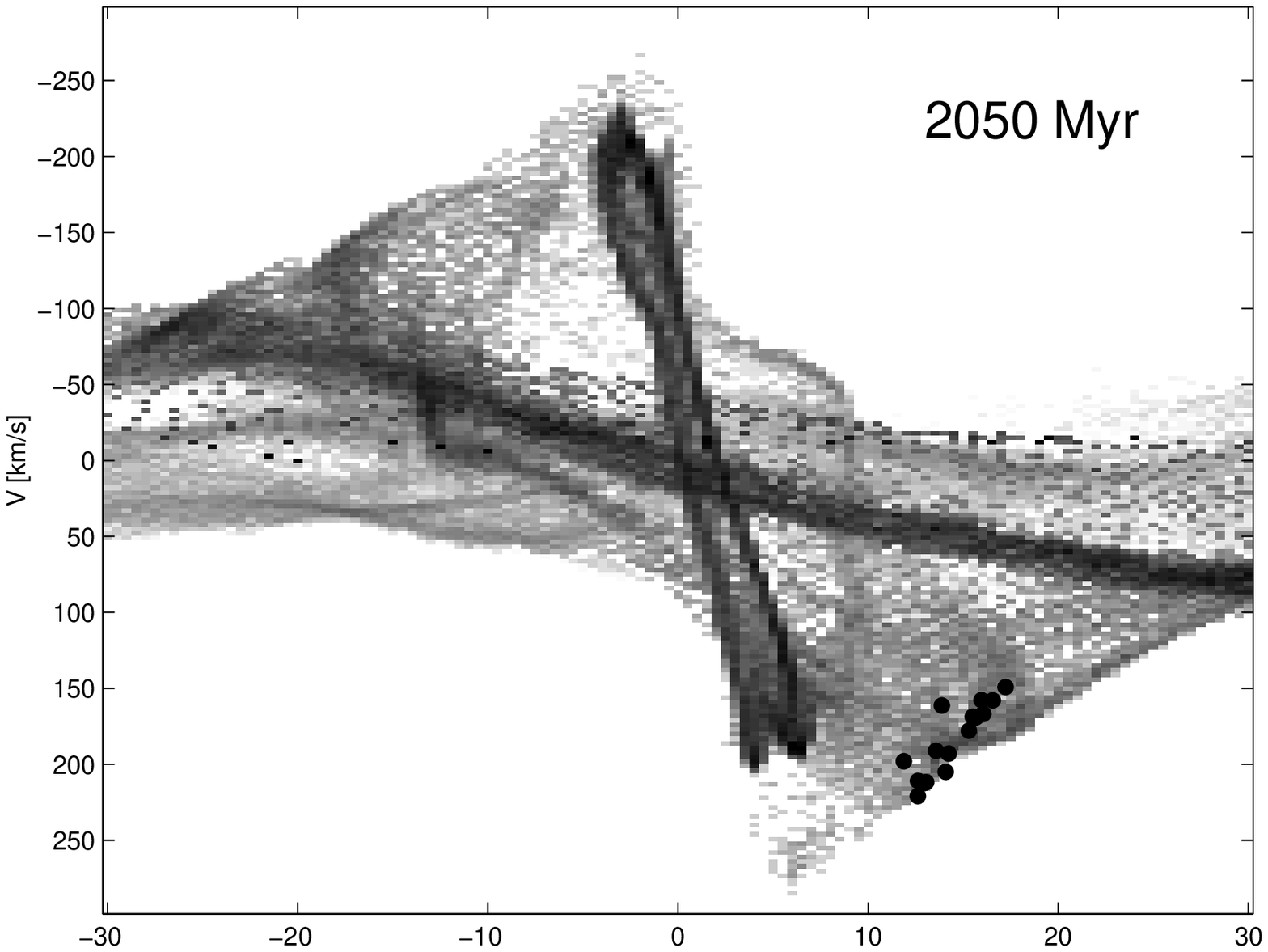,width=6.025cm}}
\centerline{\epsfig{file=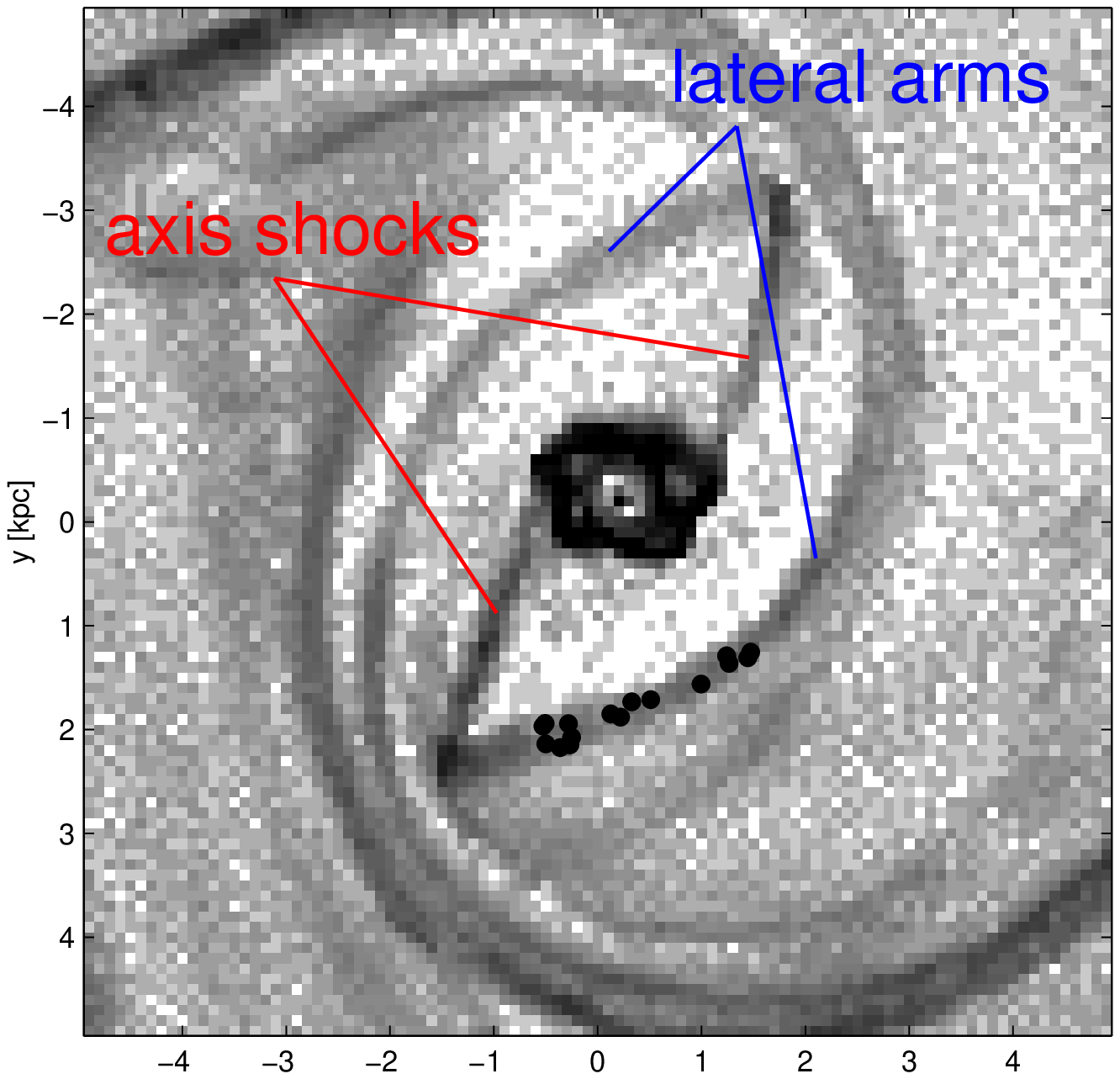,width=5.5cm}
            \epsfig{file=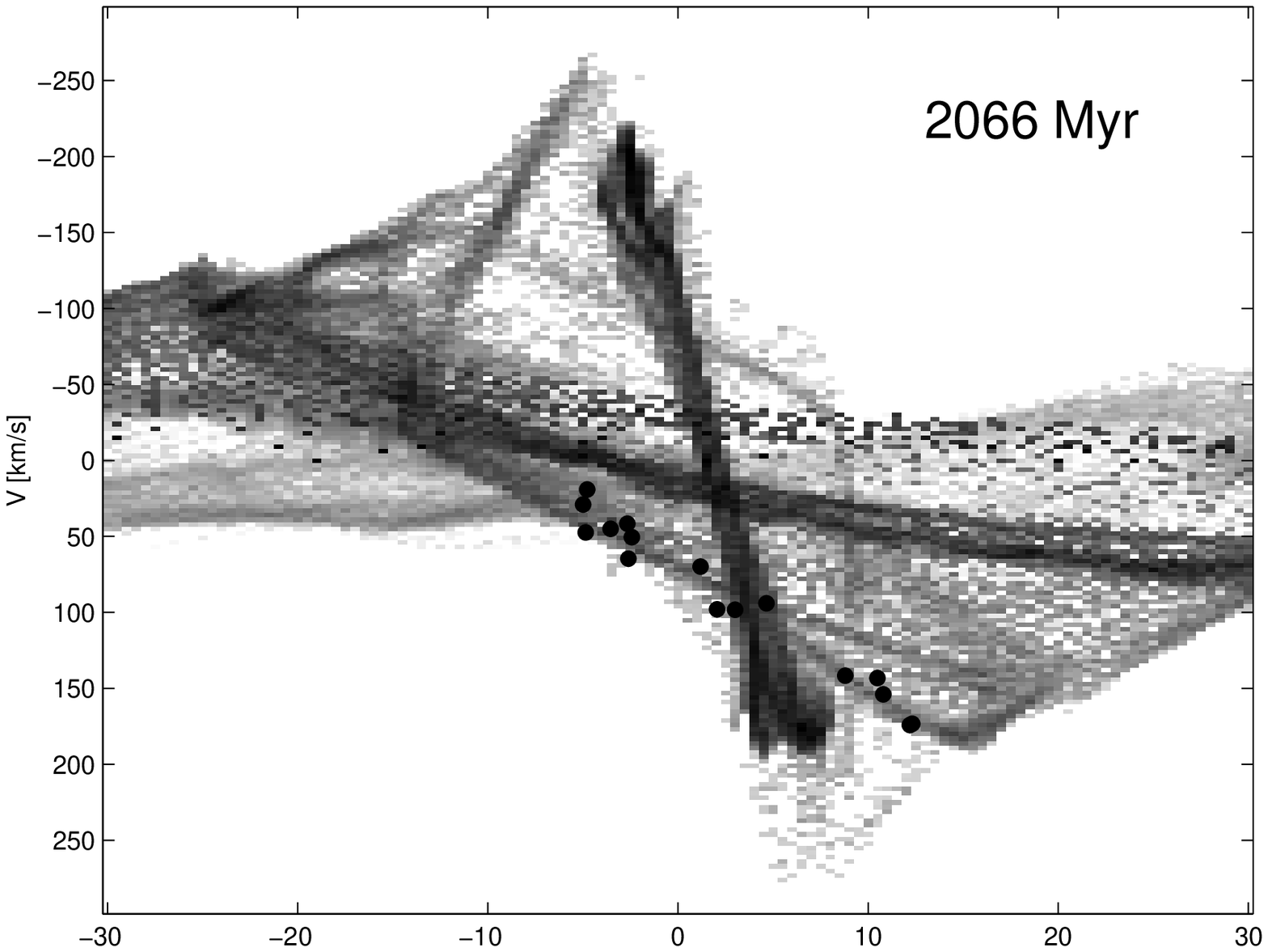,width=6.025cm}}
\centerline{\epsfig{file=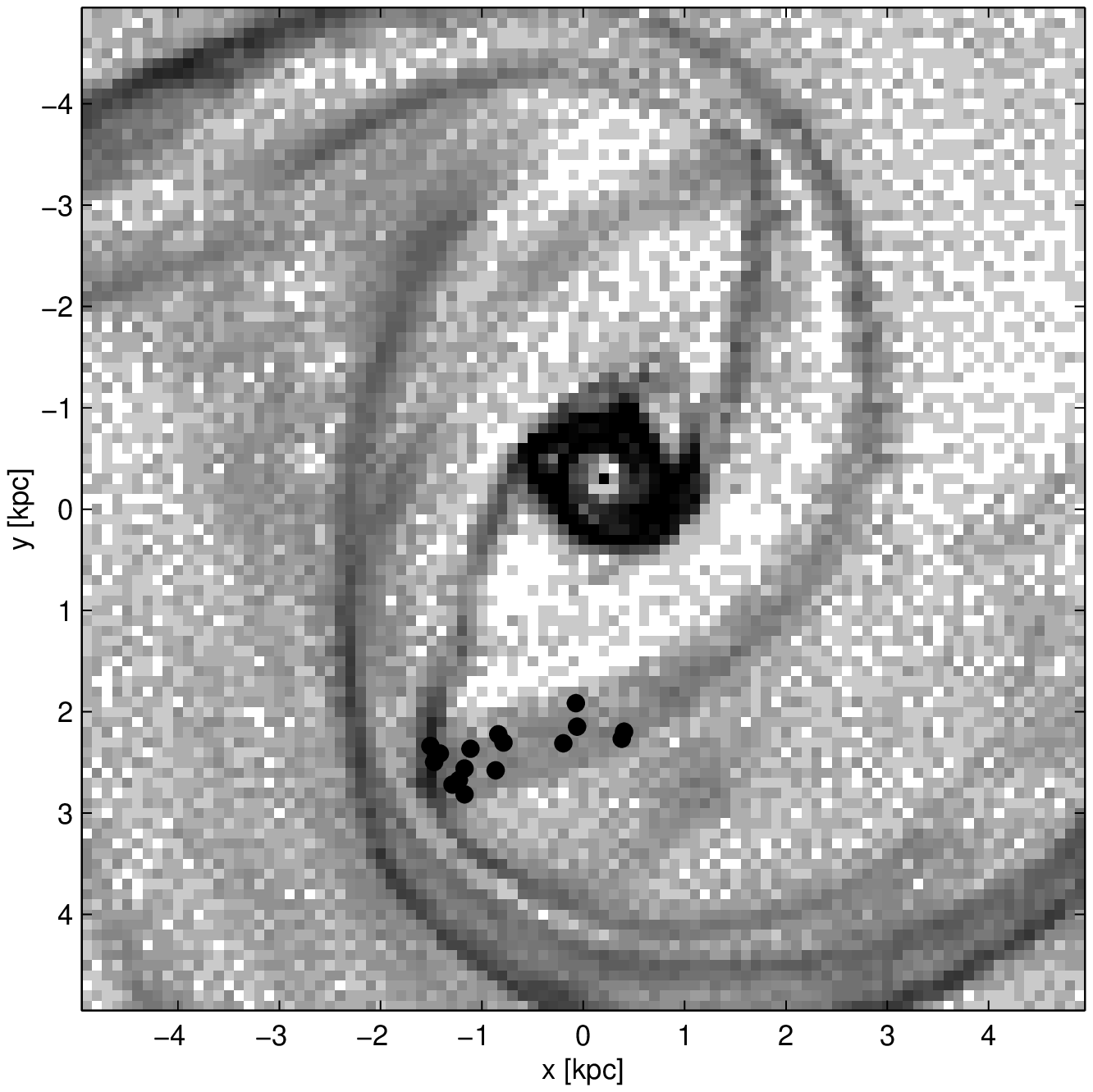,width=5.5cm}
            \epsfig{file=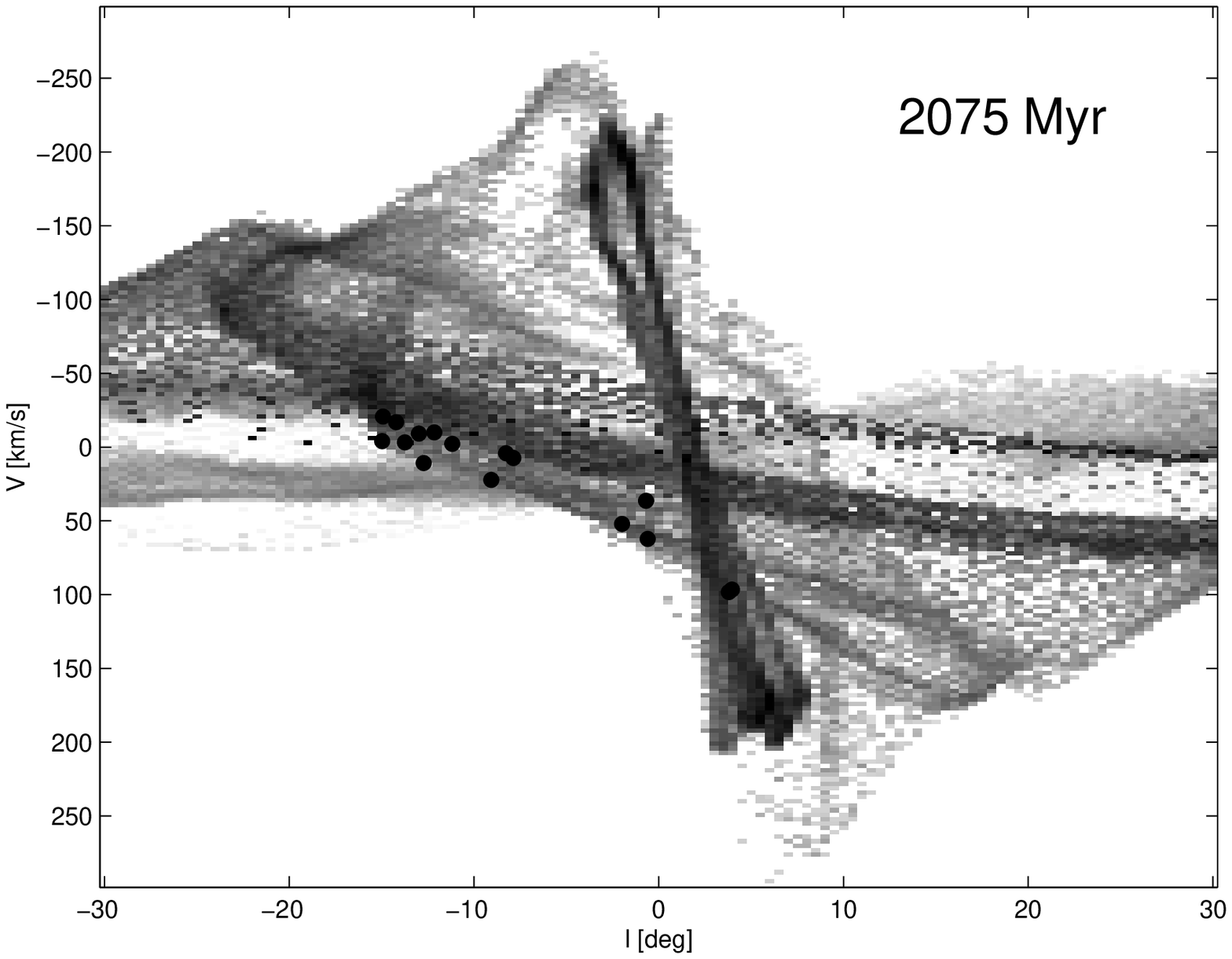,width=6.025cm}}
\caption{{\it See next page}.}
\end{figure}
\addtocounter{figure}{-1}
\begin{figure}[t!]
\centerline{\epsfig{file=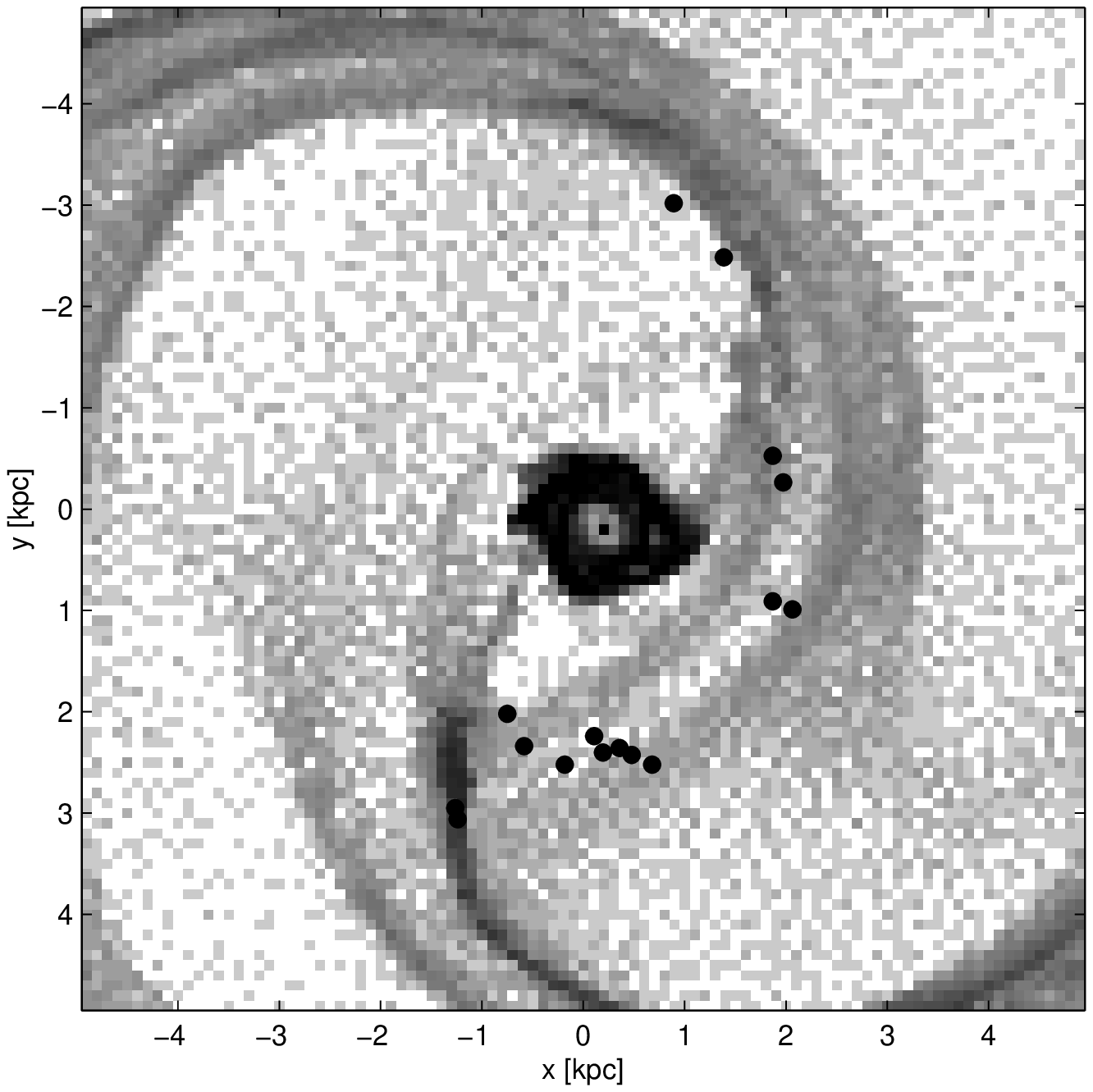,width=5.5cm}
            \epsfig{file=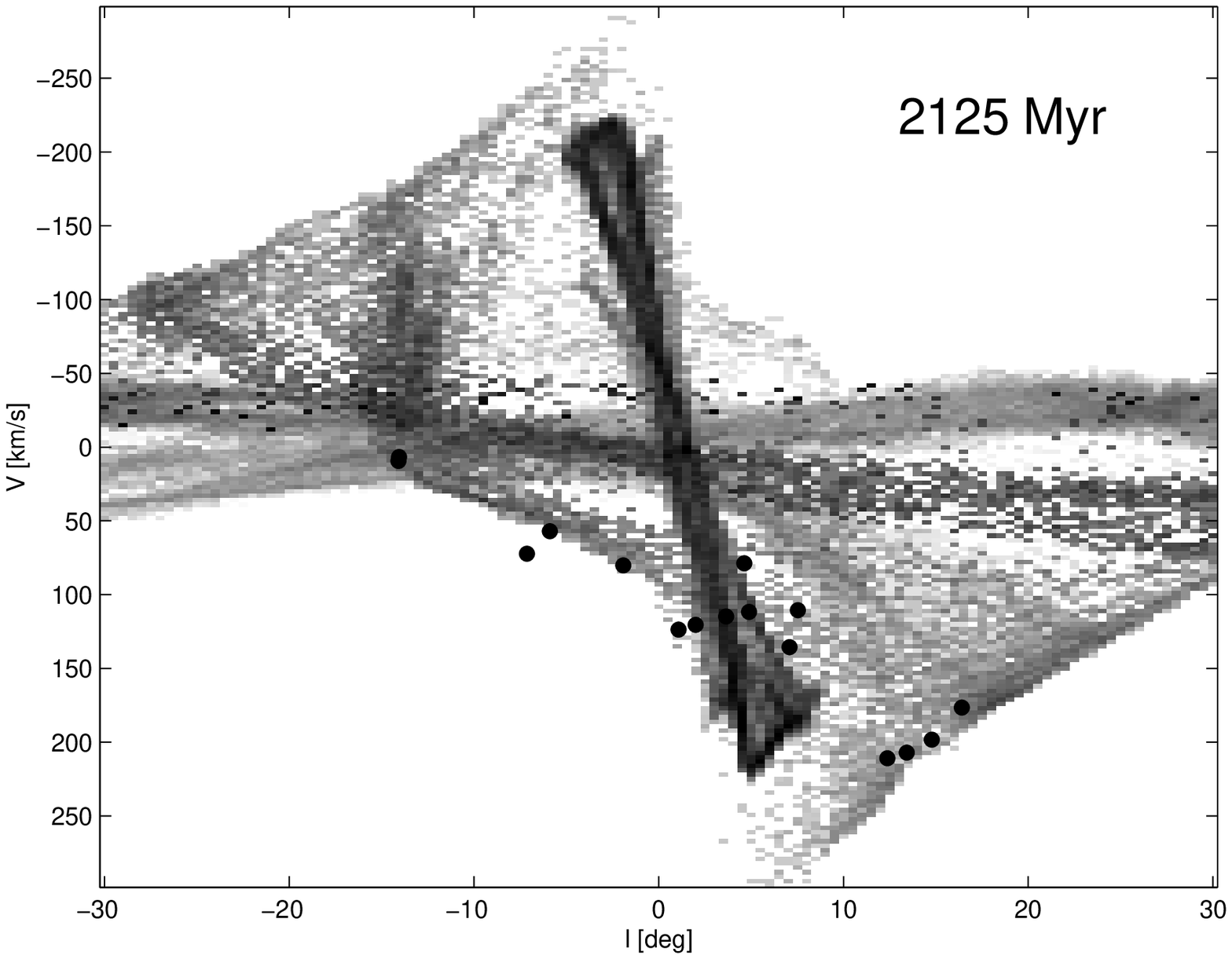,width=6.025cm}}
\caption{{\it Continued}. In simulation l10, trajectories of
collisionless particles with phase space coordinates similar to the
3-kpc arm gas at $t=2066$~Myr, i.e. located within 100~pc and
14~km\,s$^{-1}$ of the gas particles in this arm having spatial
densities over 0.1~$M_{\odot}$\,pc$^{-3}$. The $x-y$ plots are in the
rotating frame of the bar and the $\ell-V$ plots for a comoving
observer such that the bar inclination angle remains at $25^{\circ}$.
In the frames at $t=2125$~Myr, the bar has been rotated by $\pi$. The
collisionless particles always remain confined inside the 3-kpc arm as
they pass from one end of the bar to the other, illustrating that the
gas in this arm follows ballistic orbits with only little dissipation
and that the matter keeps locked in the arm.}
\label{traj}
\end{figure}
The 3-kpc and the 135-km\,s$^{-1}$ arms are lateral arms, with the
former arm ending in the near-side axis shock. They are the inner
prolongations of spiral arms in the disc and their velocity asymmetry
at $\ell=0$ comes from the fact that the 135-km\,s$^{-1}$ arm passes
much closer and faster to the Galactic centre.
\par Sevenster (1997) has outlined 9 OH/IR stars following very
closely the 3-kpc arm pattern. From their outflow velocities and
assuming disc metallicities, she infers masses between 3.5 and
6~$M_{\odot}$ and thus ages between 100 and 350~Myr for these stars,
corresponding to a few galactic rotations. Since these stars are still
coincident with a gaseous density maximum, Sevenster (1999) concludes
that the 3-kpc arm cannot be a density wave, but rather is part of an
inner ring as observed in many external barred galaxies (Buta 1996).
\par In our simulations, the lateral arms cannot be considered as true
density waves. Indeed, in the rotating frame of the bar, gas on the
lateral arms moves in a direction almost parallel to the maximum
density line of the arms and thus there is no net propagation of them
relative to the sustaining medium. Consequently, comoving stars may
be captured inside such an arm longer than if the flow were
perpendicular to the arm. Figure~\ref{traj} shows the trajectories of
collisionless particles whose phase space coordinates are similar to
the gaseous ``3-kpc'' arm in model l10t2066. At $t=2025$~Myr, these
particles are all concentrated at the far end of the bar, where they
are almost corotating with the bar. Referring to the enhanced star
formation rate observed at the bar ends of several external galaxies
(e.g. Reynaud \& Downes 1998 for NGC~1530), which may result from the
compression of the gas as lateral arms strike the outer part of the
axis shocks, this concentration may suggest that the OH/IR stars
tracing the 3-kpc arm have formed in this region. If this is correct,
then these OH/IR stars should be $\sim 40$~Myr old and thus more
massive than 6~$M_{\odot}$ (in the simulations, the particles tracing
the \mbox{3-kpc} arm turn round the bar in
$2\pi/(\Omega-\Omega_P)\approx 120$~Myr). It is also possible that
these OH/IR stars formed earlier at the near end of the bar, made more
than half a rotation round the bar on quasi-periodic $x_1$ orbits and
finally mixed with the 3-kpc arm before passing again through their
place of birth (see $t=2125$~Myr in Figure~\ref{traj}). In this case,
the age of the OH/IR stars would be $\sim 100$~Myr, becoming
consistent with Sevenster's (1997) estimation. In principle, this
interpretation can even be generalised to several pre-rotations of the
stars before they are observed within the 3-kpc arm, though phase
mixing will spread the stars around the bar.

\begin{acknowledgements}
The author thanks L.~Martinet, D.~Pfenniger, D.~Friedli, L.~Blitz,
F.~Combes, P.~Englmaier, O.~Gerhard and J.~Sellwood for suggestive
discussions, and D.~Pfenniger, D.~Friedli and W.~Benz for the original
version of the $N$-body/SPH code.
\end{acknowledgements}

\end{article}

\end{document}